\documentclass[aps,pre,10pt,showpacs,nofootinbib,twocolumn,superscriptaddress]{revtex4-1}

\usepackage{amsmath}
\usepackage{amsfonts}
\usepackage{amssymb}
\usepackage{graphicx}
\usepackage{subfigure}
\usepackage{hyperref}
\usepackage{color}
\usepackage[usenames,dvipsnames]{xcolor}
\usepackage[latin1]{inputenc}
\usepackage{comment}

\begin{document}

\definecolor{gray}{rgb}{0.4,0.4,0.4} 
\newcommand{\red}[1]{{\color{red}{#1}}}
\newcommand{\blue}[1]{{\color{blue}{#1}}}
\newcommand{\km}{\langle k \rangle}
\newcommand{\ksq}{\langle k^2 \rangle}
\newcommand{\rhot}{\varrho}
\newcommand{\Note}[1]{{\color{red}\sc ({#1})}}
\newcommand{\NoteS}[1]{{\color{blue}\sc (SCF: {#1})}}
\newcommand{\avk}[1]{\langle {#1} \rangle}


\newcommand{\FigPath}{./}

\title{Effects of local population structure in a reaction-diffusion
model of a contact process on metapopulation networks}

\author{Ang\'elica S. Mata} \author{Silvio C. Ferreira}
\affiliation{Departamento de F\'{\i}sica, Universidade Federal de
  Vi\c{c}osa, 36571-000, Vi\c{c}osa - MG, Brazil}

\author{Romualdo Pastor-Satorras}
 
\affiliation{Departament de F\'{\i}sica i Enginyeria Nuclear,
  Universitat Polit\`ecnica de Catalunya, Campus Nord B4, 08034
  Barcelona, Spain}

\begin{abstract}
  We investigate the effects of local population structure in
  reaction-diffusion processes representing a contact process (CP) on
  metapopulations represented as complex networks.  Considering a
  model in which the nodes of a large scale network represent local
  populations defined in terms of a homogeneous graph, we show by
  means of extensive numerical simulations that the critical
  properties of the reaction-diffusion system are independent of the
  local population structure, even when this one is given by a ordered
  linear chain. This independence is confirmed by the perfect matching
  between numerical critical exponents and the results from a
  heterogeneous mean field theory suited, in principle, to describe
  situations of local homogeneous mixing.  The analysis of several
  variations of the reaction-diffusion process allow to conclude the
  independence from population structure of the critical properties of
  CP-like models on metapopulations, and thus of the universality of
  the reaction-diffusion description of this kind of models.
\end{abstract}

\pacs{89.75.Hc, 05.70.Jk, 05.10.Gg, 64.60.an}

\maketitle

\section{Introduction}
\label{sec:intro}
The interplay between topology and dynamics rates among the most
outstanding challenges in complex network
theory~\cite{newman2010}. Supported by a large amount of scientific
evidence, it has become clear that dynamical processes running on top
of a complex network can be heavily influenced by the topological
properties of the substrate \cite{barratbook,dorogovtsev07}, specially
when the network shows a strongly heterogeneous pattern of
connections, as in the case of the so-called scale-free (SF) networks,
with a degree distribution (probability that a vertex is connected to
$k$ others) scaling as a power-law $P(k) \sim k^{-\gamma}$
\cite{Barabasi:1999}. Initial research on particle interaction models
in networks focused on \textit{fermionic} models, in which every node
of the network can be occupied by at most one particle. Within this
framework, a number of remarkable results were developed, concerning
the effects of large degree fluctuations in the behavior of processes
both in and out of equilibrium
\cite{leoneising,PhysRevE.66.016104,Cohen00,PhysRevLett.85.5468,Romu,newman02,
  Lloyd18052001,Castellano:2006}.

Recently, a new theoretical framework has been proposed to study
general dynamical processes on complex networks, based in the concept
of reaction-diffusion processes on metapopulations
\cite{Colizza07,baronchelli08,Ferreira07}.  Reaction-diffusion (RD) processes
\cite{vankampen} are dynamical systems defined in terms of different
kinds of particles or ``species,'' which diffuse stochastically and
interact among them according to a given set of reaction rules.  On
the other hand, metapopulation is a concept coined in ecology that
refers to the structural organization of populations as discrete
entities that interact via migration or gene flow~\cite{Hanski01}.
Metapopulation ecology deals, among other subjects, with the local
extinction and recolonization of spatially isolated patches, the
evolutionary and genetic events involved in these process as well as
the impact of human activity in the population
dynamics~\cite{Hanski01}.

In the context of network theory, considering RD processes on
metapopulations allows a \textit{bosonic} view, in which a node can
host more than one particle, particle interactions take place inside
the nodes of the network (thus representing the metapopulations), and
particle transport (mediated by diffusion) takes place among nodes
connected by edges.  RD processes on metapopulations have been
successfully applied to study different dynamics
\cite{baronchelli08,Nakao:2010fk}, but their most fruitful application
is on epidemic
spreading~\cite{Colizza08,Saldana,Balcan09,Barthelemy10}, where
metapopulations provide a realistic description of the multi-scale
socio-geographical organization of people in countries, cities,
neighborhoods, and so forth, where epidemics takes place, and in which
the transportation patterns are, in general, highly
complex~\cite{Balcan11}.

An important common factor in most previous models of RD processes on
metapopulations is that, while they take into account the
heterogeneity of the connectivity pattern of the set of populations
(in terms of a complex network), they usually disregard the internal
structure of the populations, taking a local homogeneous mixing
approximation, equivalent to describing at mean-field level the
reactions taking place inside each node. This approach raises the
issue of the possible effects of the population inner structure in the
behavior of the RD processes In this paper, we tackle this problem by
considering a metapopulation RD version of a contact process (CP)
\cite{marrobook} on a network in which each vertex represents a
structured population of fixed size given by a graph (see
Fig.~\ref{fig:model}).  By means of extensive numerical simulations,
we show that the critical properties of the RD process are insensitive
to the population structure, even in the extreme case of non-small
world \cite{watts98} linear chains. This observation is confirmed by
the perfect matching of numerical results with a heterogeneous
mean-field solution \cite{dorogovtsev07,barratbook}, that completely
neglects population structure The analysis of several variations of
the model lead us to the extended conclusion that that critical
properties of CP-like models on metapopulations depend exclusively on
the topological structure of the metapopulation network.

We have organized the paper as follows: The reaction-diffusion process
considered is defined in Section~\ref{sec:model}, and the
corresponding HMF theory is developed in Sec.~\ref{sec:hmf}.  The
results of HMF theory are compared with quasi-stationary numerical
simulations in Section~\ref{sec:simu}.  Finally, we draw our
concluding remarks in Sec.~\ref{sec:conclu}. 

\section{RD Model definition}
\label{sec:model}

\begin{figure}[t]
 \centering
 \includegraphics[width=8cm]{\FigPath/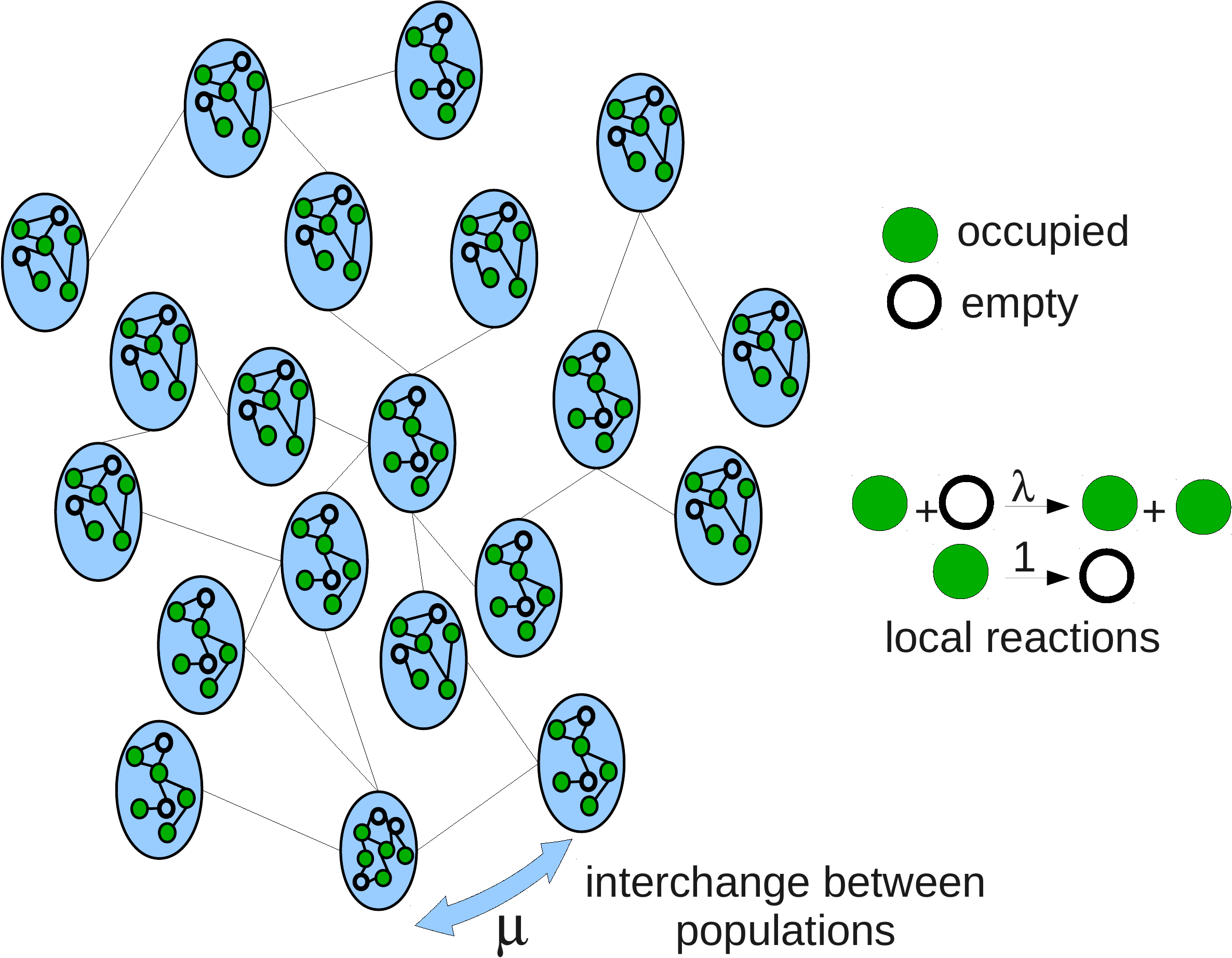}
 \caption{(color online) Reaction-diffusion process in heterogeneous
   metapopulations represent with a networked inner structure. 
  Populations are depicted by large blue vertices while the vertices inside
each population are represented by circles. Particles in connected 
   populations are exchanged at a rate $\mu$, create offspring in
   their nearest empty neighbors inside their respective
   populations at a rate $\lambda$,  and become empty at a unitary rate.}
 \label{fig:model}
\end{figure}

We consider a RD process on metapopulations representing a contact
process characterized by two typical scales: local populations where
reaction processes of creation and annihilation take place, and
non-local (long-range) interconnections between populations, see
Fig.~\ref{fig:model}.
%
The system is defined in terms of $N$ populations (the metapopulation)
of fixed size $L$ connected by non-directed edges, defining a
non-local network with a degree distribution that we choose to have a
SF form, $P(k) \sim k^{-\gamma}$. Exchange of particles, via
diffusion, takes place along the edges of this network.  The
populations themselves are considered as homogeneous graphs with
average degree $z$.  We have considered two forms for the local
population network, a random regular network and a linear chain. The
dynamics is defined as follows: Vertices inside populations may be in
two states, either occupied or empty. An occupied site creates
offspring in all empty nearest neighbors (NNs) inside its population
at a rate $\lambda$; occupied sites, on the other hand, annihilate
spontaneously at rate $1$ (this choice arbitrarily defines the
dynamics time scale). Finally, particles migrate between two connected
populations at a rate $\mu$. A migration event is performed by
swapping a particle in a population and a vertex in a population
connected by an edge.  In our model only occupied sites migrate.

The state devoid of particles represents an absorbing configuration
from which the dynamics cannot escape. We therefore expect that the system
undergoes an absorbing-state phase transition \cite{marrobook},
characterized by the order parameter $\rho$, defined as the average
density of particles. The transition will take place at some critical
point $\lambda_c(\mu)$, separating an active phase, with a constant
average density of particles ($\rho >0$), and the absorbing state with
$\rho=0$.


\section{HMF theory}
\label{sec:hmf}

The analytical study of the absorbing-state phase transition in this
model is a complex task, particularly in the case of local population
in the shape of a linear chain. We will therefore make the strong
simplification of assuming 
homogeneous mixing inside populations. In this case, the problem is
amenable to analytical solution by applying the heterogeneous
mean-field (HMF) approximation \cite{barratbook,dorogovtsev07}. Under
these conditions, the evolution equation for the density $\rho_k$ of
occupied sites in a population of degree $k$ takes the form
\begin{eqnarray}
 \frac{d\rho_k}{dt} & = & -\rho_k+z\lambda \rho_k(1-\rho_k)+
\mu k (1-\rho_k)\sum_{k'}\frac{\rho_{k'}P(k'|k)}{k'} \nonumber \\ 
& & - \rho_k\mu\sum_{k'}(1-\rho_{k'})P(k'|k),
\label{eq:rhok1}
\end{eqnarray} 
where the conditional probability $P(k'|k)$ gives the probability that
a node of degree $k$ is connected to a node of degree $k'$
\cite{alexei}.  The first term of Eq. \eqref{eq:rhok1} represents
spontaneous annihilation, while the second one represents a creation
event in empty sites with $z$ NNs.  Notice that these two terms assume
homogeneous mixing.  The third and fourth terms represent the incoming
and outgoing diffusive flow contributions due to migration of occupied
sites.  The factor $1/k'$ in the third term reckons particles
migrating to neighbor populations with equal chance.  A stability
analysis of Eq.~(\ref{eq:rhok1}) around the trivial empty state fixed
point $\rho_k=0$ yields the linearized equation:
 \begin{equation}
  \frac{d\rho_k}{dt} = \sum_{k'}L_{kk'}\rho_{k'}(t)+\mathcal{O}(\rho_k^2),
\end{equation}
where the Jacobian matrix is
$L_{kk'}=-(1-z\lambda+\mu)\delta_{kk'}+\mu k P(k'|k)/k'$,
$\delta_{kk'}$ being the Kronecker delta symbol. The active phase is
obtained when the trivial fixed point becomes unstable, which happens
if the largest eigenvalue of the Jacobian is positive.  Obviously, the
largest eigenvalue of $L_{kk'}$ can be written as
$\ell_m=-(1-z\lambda+\mu)+\mu c_m$, where $c_m$ is the largest
eigenvalue of $C_{kk'}=k P(k'|k)/k'$. It is easy to show that $v_k=k$
is an eigenvector of $C_{kk'}$ with eigenvalue $c=1$.  The matrix
$C_{kk'}$ is irreducible because the network is connected and,
consequently, Perron-Frobenius theorem~\cite{newman2010} implies that
$C_{kk'}$ has a unique eigenvector with positive components that
corresponds to the largest eigenvalue.  Therefore, $c_m=c=1$ and
condition $\ell_m>0$, gives the critical point $\lambda_c=1/z$.

To obtain information about other critical properties, we consider
uncorrelated networks for which $P(k'|k)=k'P(k')/\km$
\cite{Dorogovtsev:2002}, which leads to the equation
\begin{equation}
  \label{eq:rhok2}
  \frac{d\rho_k}{dt}  = z\Delta \rho_k -\lambda z \rho_k^2+ \frac{\mu k
    \rho}{\langle k \rangle} (1-\rho_k) -
  \rho_k\mu\left(1-\frac{\Theta}{\km}\right), 
\end{equation}
where $\Delta = \lambda-\lambda_c$, $\rho = \sum_k\rho_k P(k)$ and
$\Theta = \sum_k k\rho_k P(k)$. From here, we can obtain the equation
for the total particle density, that takes the form
\begin{equation}
\label{eq:drho_onesite}
\frac{d\rho}{dt} = -\rho+z\lambda(\rho-\langle\rho_k^2\rangle),
\end{equation}
where the diffusion terms have cancelled due to conservation of
particles implied in this process.
From this expression we obtain in the steady state
\begin{equation}
 \langle\rho_k^2 \rangle  =  \Delta\rho/\lambda,
\label{eq:rhok_sqr}
\end{equation}
where $\langle\rho_k^2 \rangle = \sum_k P(k) \rho_k^2$. 

A quasi-static approximation \cite{michelediffusion},
assuming $\dot{\rho}_k\approx 0$ and isolating $\rho_k$ in
Eq.~(\ref{eq:rhok2}), yields
\begin{equation}
 \label{eq:rhok3}
 \rho_k = \frac{\rhot k}{1+\rhot k}+\frac{(\rhot k)^2}{(1+\rhot
   k)^3}+\mathcal{O}[(\rhot k)^3], 
\end{equation}
where $\rhot = \rho/[\km(1-z\Delta/\mu)-\Theta]$, and we have
performed an expansion up to order $\rho^2$, which should be valid
very close to the transition. Inserting Eq.~(\ref{eq:rhok3}) into the
definition of $\avk{\rho_k^2}$, replacing the summation by an integral
for a normalized degree distribution $P(k)=A_\gamma k^{-\gamma}$ with
$k=k_0,\cdots, k_c$, $A_\gamma=(\gamma-1)k_0^{\gamma-1}$, $k_c\gg k_0$
and $\gamma>2$, we obtain, in the limit $\rhot k_c\gg 1$,
corresponding to a super-critical phase in a infinite system limit,
\begin{equation}
\label{eq:rhok2_asym}
\langle\rho_k^2 \rangle  =  \left\lbrace 
 \begin{array}{lll}
   \dfrac{(\gamma-1)\Gamma(3-\gamma)\Gamma(\gamma)}{2}(k_0\rhot)^{\gamma-1} & , & 2<\gamma<3 \\
  \dfrac{(\gamma-1)^2}{\gamma-3} (k_0 \rhot)^2&, & \gamma>3
   \end{array}
 \right.,
\end{equation}
where $\Gamma(x)$ is the Gamma function~\cite{gradshteyn2007}.  

From Eqs.~(\ref{eq:rhok_sqr}) and (\ref{eq:rhok2_asym}), and using
that $\rhot \simeq \rho/\km$ very close to critical point, we find
$\rho\sim\Delta^\beta$, with a critical exponent
$\beta={1}/({\gamma-2)}$ if $2<\gamma<3$ and $\beta=1$ if $\gamma\ge
3$.  At criticality, setting $\lambda = 1/z$, the density evolves as
$d\rho/dt = -\langle \rho_k^2\rangle$.  Using the asymptotic form of
$\langle\rho_k^2 \rangle$ given by Eq.~(\ref{eq:rhok2_asym}), one
obtains $\rho\sim t^{-\delta}$ with $\delta = \beta$. We can also show
that close to the critical point the density approaches the asymptotic
value as $\rho(t)=\rho_s+\mbox{const.} \times \exp(-t/\tau)$ where the
characteristic time diverges as $\tau=(zb\Delta)^{-1}\sim
\Delta^{-\nu_\parallel}$ with an exponent $\nu_\parallel=1$.

Interestingly, these critical exponents coincide with the HMF solution
for the contact process in uncorrelated networks
\cite{castellano08,boguna09}. This fact suggests an extension of the
contact process approach to compute the finite-size scaling (FSS)
behavior of our model. Indeed, applying the strategy proposed in
Ref.~\cite{castellano08}, in which the motion equation is mapped in a
one-step process, we find (see Appendix~\ref{app:fss} for details)
that the stationary density of occupied vertices, $\bar{\rho}$, and
the characteristic time, $\tau$, display  an
anomalous size dependence at the transition given by
\begin{equation}
  \bar{\rho} \sim \left(\Omega g \right)^{-1/2} \mbox{~and~}
  {\tau}\sim \left(\Omega/g\right)^{1/2}.
  \label{eq:FSS}
\end{equation}
where $g=\frac{\ksq}{\km^2}$ and $\Omega=N\cdot L$.  Since $g=\ksq/\km^2$ depends only of $N$
through the network cut-off $k_c \sim N^{1/\omega}$ \cite{Boguna2004},
we have that $\bar{\rho}\sim L^{-1/2}N^{-\hat{\nu}}$ and $\tau\sim
L^{1/2}N^{\hat{\alpha}}$ where
\begin{equation}
  \hat{\nu} = \frac{1}{2}+\max\left(\frac{3-\gamma}{2\omega},0\right),~~
  \hat{\alpha} = \frac{1}{2}-\max\left(\frac{\gamma-3}{2\omega},0\right).
  \label{eq:exps}
\end{equation}

\section{Quasi-stationary simulations}
\label{sec:simu}

\subsection{Methods}

In order to check the predictions of the HMF theory we performed
extensive simulations of the RD process in heterogeneous
metapopulations.  The networks connecting populations were
simulated using the uncorrelated configuration model
(UCM)~\cite{Catanzaro05}, that warrants the absence of degree
correlation.  Letting $(i,j)$ represent the vertex $j$ inside the
population $i$, the computer algorithm is implemented as follows: At
each time step, an occupied vertex $(i,j)$ is chosen and the time
increased by $\Delta t = 1/[n(1+\lambda+\mu)]$, where $n$ is the
number of particles at time $t$. With probability
$p=1/(1+\lambda+\mu)$, the vertex becomes vacant. With probability
$q=\lambda/(1+\lambda+\mu)$, all empty nearest neighbors of $j$ inside
population $i$ are occupied.  Finally, the migration to a randomly
chosen vertex $(i',j')$ belonging to a neighbor population occurs with
probability $r=1-p-q$.  

We used a quasi-stationary (QS) simulation method where, every time the system visits the
absorbing state, the absorbing configuration is replaced by a state
randomly taken from the system history~\cite{DeOliveira05,Ferreira11a}. To
implement the method, a list containing $M=100$ configurations is
stored and constantly updated. The updating is done by randomly
picking up a stored configuration and replacing it by the current one
with probability $p_r\Delta t$.  We fixed $p_r\simeq 10^{-4}-10^{-3}$
(the larger $N$ the smaller $p_r$). After a relaxation time $t_r =
10^6$, the QS averages are computed over times up to $t_{av}=2\times
10^7$. Ensembles of 50 network configurations were used for
statistical averages.

\begin{figure}[hbt]
  \centering
 \includegraphics[width=8cm]{\FigPath/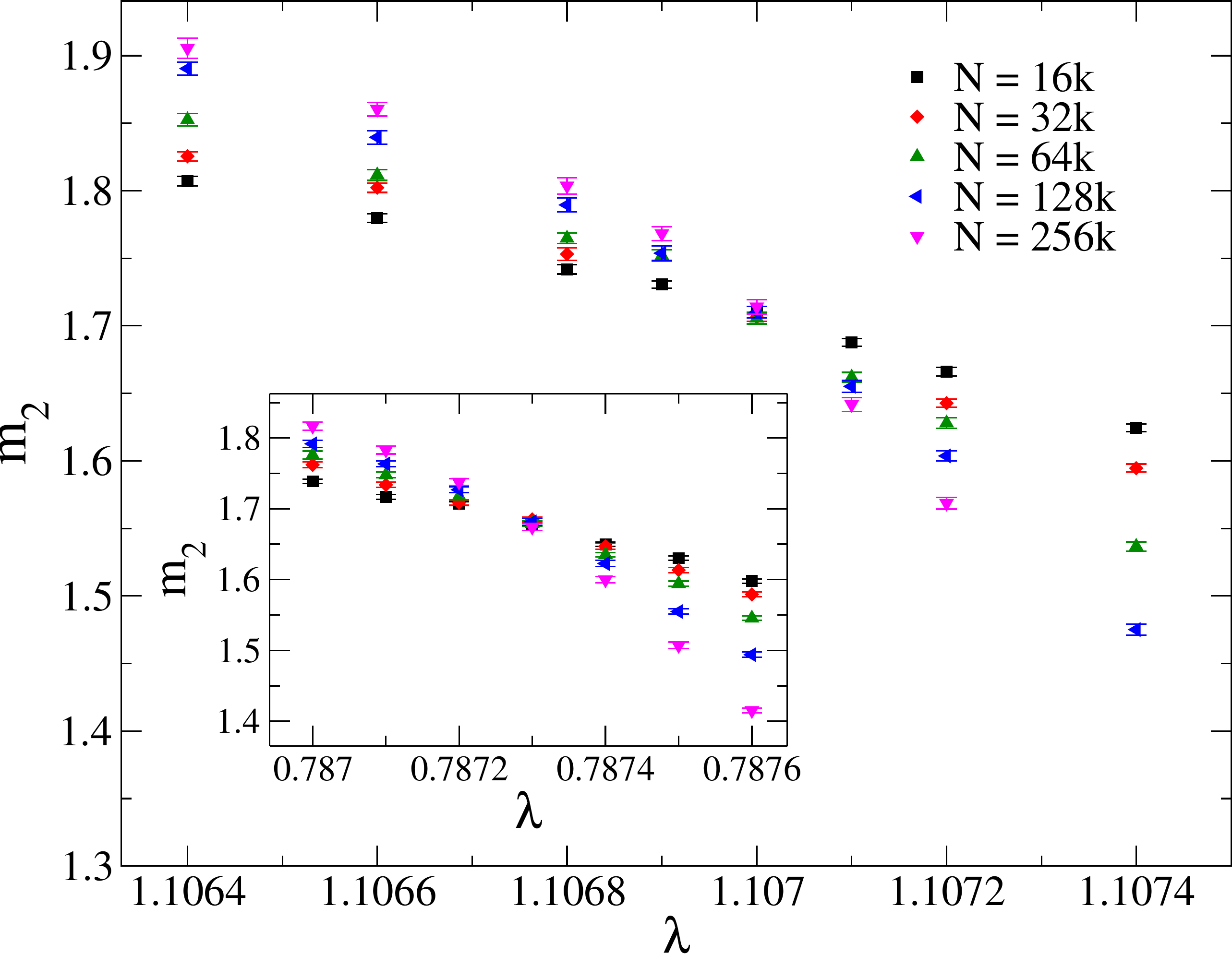} 


  \caption{ \label{fig:moments} Moment ratios for the
    reaction-diffusion processes on metapopulations connected by UCM
    networks with minimum degree $k_0=6$ and cutoff $k_c =
    N^{1/2}$. In the main plot, the migration rate is $\mu=0.10$,
    degree exponent is $\gamma=2.75$ and the local population
    structure of the populations are RRNs with size $L=250$ and
    $z=2$. In the inset, the parameters $\mu=0.50$,
    $\gamma=2.25$ and regular chains of size $L=250$ and $z=2$ are
    used.}
 \end{figure}

The method used to determine critical points in simulations is based
in the idea of the size independence at criticality of the moment
ratio $m_2 = \langle \rho^2\rangle/\langle\rho\rangle^2$
\cite{Ferreira11a,Ferreira11b,sander_phase_2013}: by plotting
the ratio $m_2$ as a function of $\lambda$ for different values of the
network size $N$, the critical point will be given by the intersection
of the plots for all network sizes.  Figure~\ref{fig:moments} shows typical curves used to determine
the critical point using the moment ratio technique.

\begin{table*}[bt]
\begin{center}
\begin{tabular}{cccccccccc}
\hline\hline
              &\multicolumn{4}{c}{Regular Random Network}  &~&\multicolumn{4}{c}{Linear Chain} \\ \cline{2-5}\cline{7-10}
              &~$\mu=$0.05~&~$\mu=$0.10~&~$\mu=$0.50~&~$\mu=$0.90 
          &~~~&~$\mu=$0.05~&~$\mu=$0.10~&~$\mu=$0.50~&~$\mu=$0.90\\ \cline{2-5}\cline{7-10}
$\gamma=$2.25  & 1.2361(1) & 1.10585(5) & 0.78995(5) & 0.68680(5) &~~~& 1.23315(5) & 1.1031(1) & 0.7873(1) & 0.69445(5)\\
$\gamma=$2.75  & 1.23705(5) & 1.1070(1) & 0.79145(5) & 0.69820(5) &~~~& 1.2336(1) & 1.1034(1) & 0.78740(5) & 0.69455(5)\\
$\gamma=$3.50  & 1.2380(1) & 1.10820(5) & 0.7930(1) & 0.69970(5) &~~~& 1.2340(5) & 1.1038(1) & 0.7876(1) & 0.69465(5)\\ 
PA             & 0.954     & 0.916      & 0.75      & 0.678      &~~~& 0.954     & 0.916     & 0.75      & 0.678\\ 
\hline\hline
\end{tabular}
 \caption{\label{tab:lambdac}
   Critical points for the RD process on populations connected by UCM
   networks with minimum degree $k_0=6$, cutoff $k_c = N^{1/2}$, and
   degree exponent $\gamma$.  The population structure consists in
   either a RRN or a linear chain with $L=250$ vertices and $z=2$. The
   thresholds in a pair approximation $\lambda_c^{PA} =
   (1+\mu)/(1+2\mu)$ are also shown. Numbers in parenthesis represent
   uncertainties in the last digit.}
\end{center}
\end{table*}

\subsection{Numerical results}

Firstly, we have considered the case of local populations with a
random structure, represented by a random regular network
(RRN)~\cite{Ferreira12} with fixed average degree $z=2$ and random
connections excluding self- and multiple
edges. Table~\ref{tab:lambdac} shows the respective critical points
for different migration rates and degree exponents.  We observe that
the critical points are almost independent of the heterogeneity of the
population network, expressed by the degree exponent $\gamma$, in
agreement with the results obtained for the contact process in a
fermionic approach~\cite{Ferreira11b}. On the other hand, we observe
that the critical point decreases for increasing migration rate, at
odds with the constant HMF prediction $\lambda_c = 1/z$. This
observation is however reasonable, since migration facilitates
activity spreading to non-active regions, and can be accounted for by
means of a simple homogeneous pair
approximation~\cite{Ferreira11b,Juhasz12,sander_phase_2013}. Within
this approach (see Appendix~\ref{app:PA}), we obtain a new threshold
$\lambda_c^{PA}(\mu) = (1+\mu)/(z-1+\mu z)$ that is closer to
simulation results than the HMF prediction.

\begin{figure}[t]
 \centering
 \includegraphics[width=8cm]{\FigPath/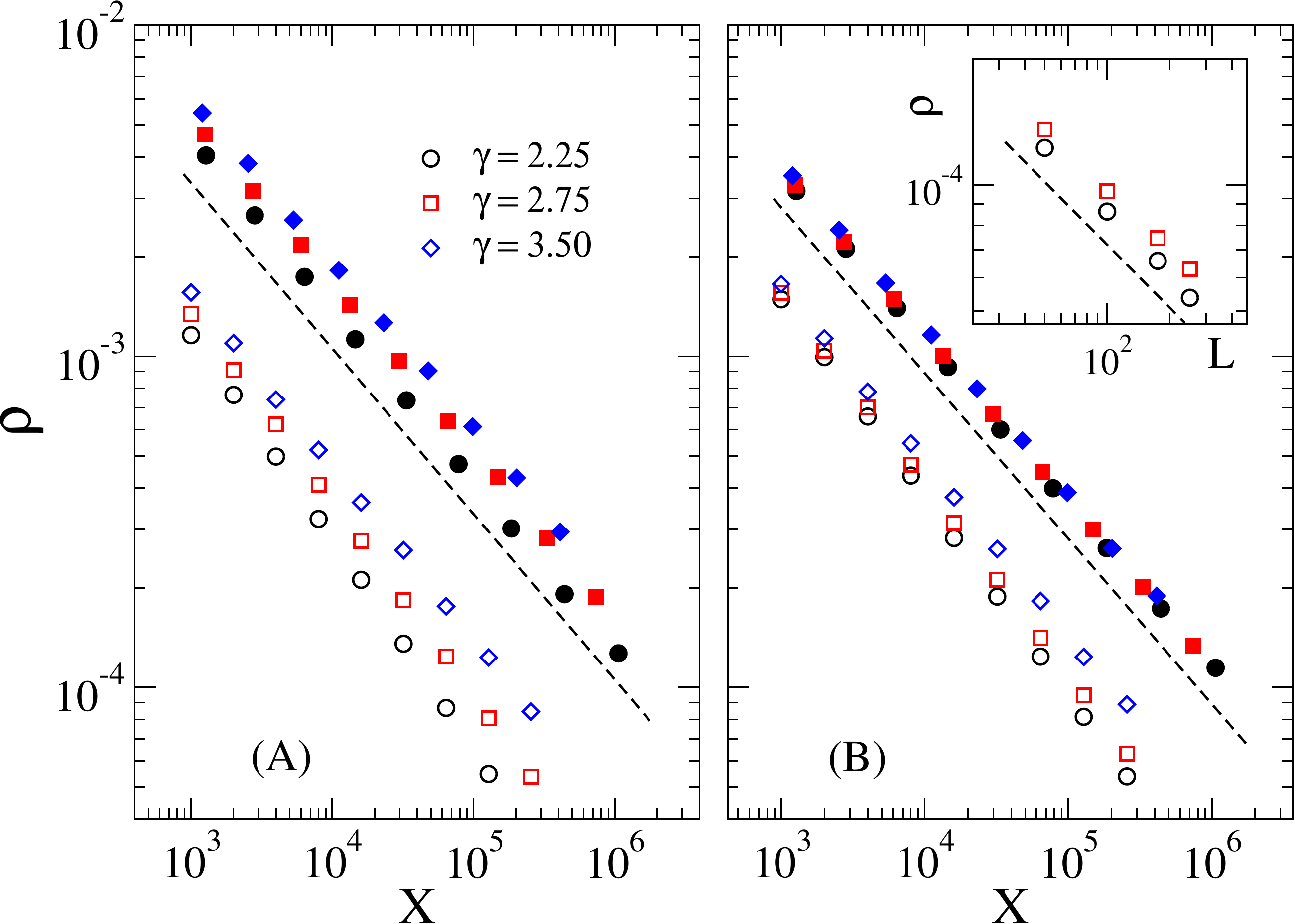}
 \caption{Finite size scaling of the critical density for distinct
   structured local populations. Results for (A) RRN and (B) linear
   chain are shown. The migration rate is $\mu=0.1$ and the remaining
   fixed parameters are as in table~\ref{tab:lambdac}. The curves
   $\bar{\rho}$ vs. $N$ are represented by open symbols while curves
   $\rho$ vs. $gN$, $g=\ksq/\km^2$, are represented by the filled
   ones. {Inset in (B) shows the critical density against $L$ for $N =
     256000$ and $\gamma=2.75$ for RRN (circles)
     and linear chains (squares). Filled symbols were shifted
     to improve visibility. Dashed lines represent the power
     law $X^{-1/2}$.}}
 \label{fig:rhoa010}
\end{figure}

\begin{figure}[ht]
 \centering
 \includegraphics[width=8cm]{\FigPath/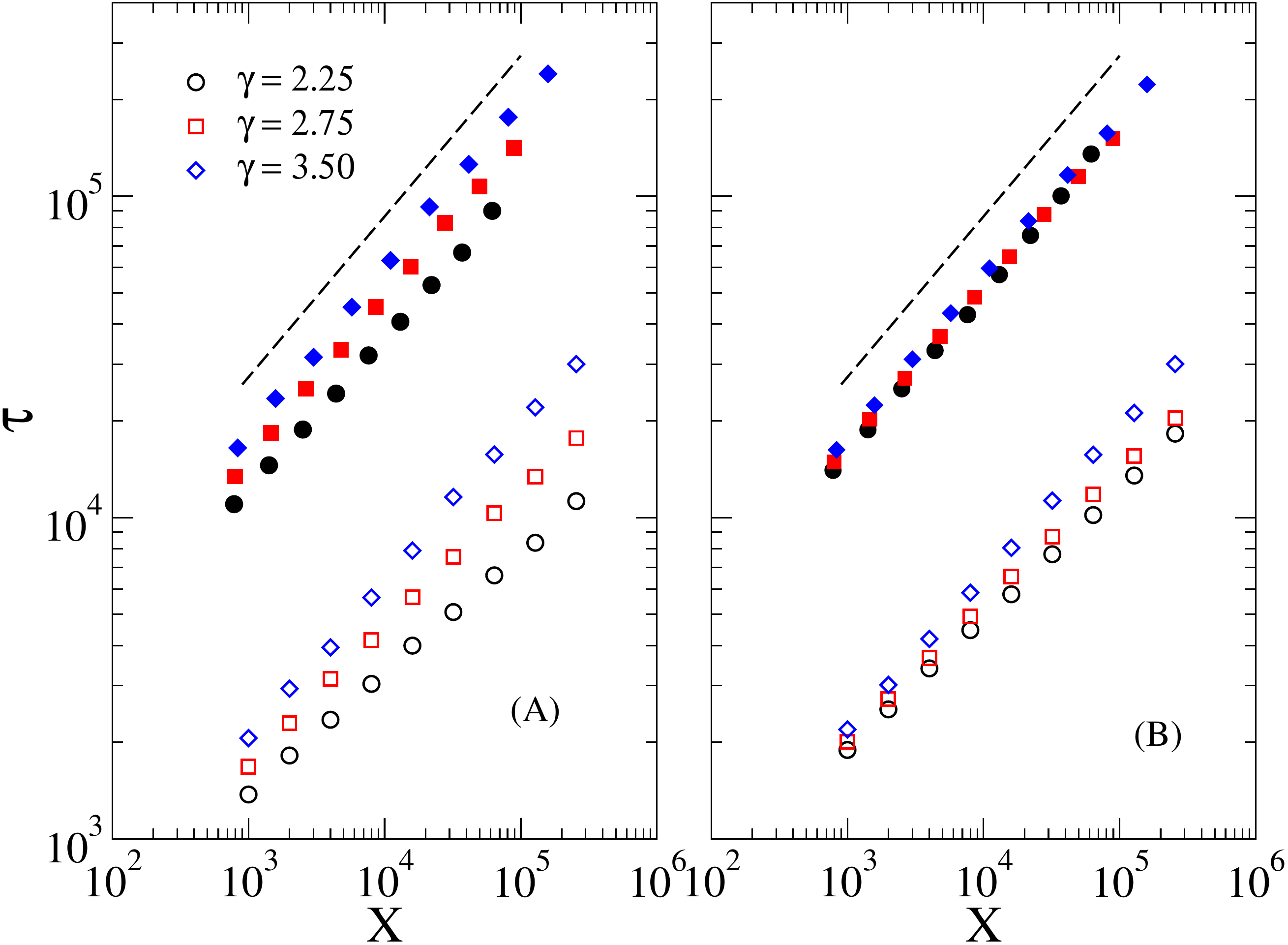} 
 \caption{ Finite size scaling of the critical characteristic
time for distinct structured local populations. Results for (A) RRN and (B)
linear chain are shown. The fixed parameters are as in
figure~\ref{fig:rhoa010}. The curves $\tau$ vs. $N$ are
represented by open symbols while curves $\tau$ vs. $N/g$, $g=\ksq/\km^2$, are
represented by the filled ones. Filled symbols were shifted to improve
visibility. Dashed lines represent the power law $X^{1/2}$.} 
\label{fig:tau_a010}
\end{figure}

The great trump of the HMF theory in absorbing state phase transitions is
to predict the correct FSS critical exponents even for quenched
networks~\cite{Ferreira11b,sander_phase_2013}, where neglecting
dynamical correlations implies an
approximation. Figure~\ref{fig:rhoa010} shows an example of FSS for
critical density $\bar{\rho}$ and migration rate $\mu=0.1$.  Similar
plots are obtained for the characteristic time $\tau$, as illustrated
in Fig.~\ref{fig:tau_a010}. The power law regressions in
plots of $\bar{\rho}$ and $\tau$ \textit{vs.} $N$ yield exponents
varying with network heterogeneity but independent of the exchange
rate. The exponents apparently differ from HMF predictions given by
Eq.~\eqref{eq:exps} as shown in Table~\ref{tab:exp}.  However, the
$\ksq$ and $\km$ factors in Eq.~(\ref{eq:FSS}) introduce strong
corrections to scaling~\cite{boguna09,Ferreira11a,Ferreira11b}, which
can be explicitly taken into account by performing power law
regressions of $\bar{\rho}$ vs. $gN$ and $\tau$ vs. $N/g$. In this
way, the HMF scaling law $\bar{\rho}\sim (gN)^{-S_\nu}$ with
$S_\nu=1/2$ is recovered in all simulations, as shown in
Fig.~\ref{fig:rhoa010} and Table~\ref{tab:exp} for $\mu=0.1$.
The scaling $\tau\sim (N/g)^{S_\alpha}$ with $S_\alpha=1/2$ is also very
well verified, as one can see in Fig.~\ref{fig:tau_a010}.  For fixed
$N$, the scaling laws $\bar{\rho}\sim L^{-1/2}$ and $\tau\sim L
^{1/2}$ are also confirmed in simulations as shown in the inset of
Fig.~\ref{fig:rhoa010}. However, the critical points slightly
increases for smaller population sizes (from $\lambda_c = 1.1070$ for
$L=250$ to $\lambda_c=1.1105$ for $L=100$).

Turning now to the more interesting case of a population with a regularly
ordered internal structure, we consider the case of a linear chain with
coordination number $z=2$. Table~\ref{tab:lambdac} shows the critical points
numerically computed for different migration rates. The critical points are
practically identical to those found for RRNs. On the other hand,
figures~\ref{fig:rhoa010} and \ref{fig:tau_a010}  shows a surprisingly 
very good quantitative agreement between FSS critical exponents of the numerical
simulations and the prediction of the HMF theory,see Table~\ref{tab:exp}.
Results
for the other values of $\mu$ yield the same critical exponent inside error
bars.

\begin{table*}[t]
\begin{center}
\begin{tabular}{ccccccccccccc}
 \hline\hline 
 &\multicolumn{4}{c}{RRN} & ~ &\multicolumn{4}{c}{Chain} & ~ &\multicolumn{2}{c}{HMF} \\  \cline{2-5}\cline{7-10}\cline{12-13}
 & $\hat{\nu}$ & $S_\nu$ & $\hat{\alpha}$ & $S_\alpha$ & &
 $\hat{\nu}$ & $S_\nu$   & $\hat{\alpha}$ & $S_\alpha$ & ~ & $\hat{\nu}$ & $\hat{\alpha}$ \\ 
\cline{2-5}\cline{7-10}\cline{12-13}
$\gamma$=2.25 ~   & 0.63(2) ~ & 0.51(2) ~ & 0.37(2) ~ & 0.48(2) ~ & &
           0.60(2) ~ & 0.49(2) ~ & 0.40(3) ~ & 0.51(2) ~ & & 0.69 ~ &  0.31\\
$\gamma$=2.75 ~   & 0.58(1) ~ & 0.49(1) ~ & 0.43(3) ~ & 0.52(3) ~ & &
           0.58(2) ~ & 0.50(2) ~ & 0.42(2) ~ & 0.49(2) ~ & & 0.56 ~ & 0.44\\
$\gamma$=3.50 ~   & 0.52(2) ~ & 0.50(1) ~ & 0.49(2) ~ & 0.51(2) ~ & &
           0.53(2) ~ & 0.50(1) ~ & 0.47(3) ~ & 0.50(3) ~  & & 1/2 ~ & 1/2 \\ \hline\hline
\end{tabular}
\caption{Critical exponents in the FSS of the quasi-stationary
  density and characteristic time: $\bar{\rho}\sim N^{-\hat{\nu}}$, $\bar{\rho}\sim
  (gN)^{-S_\nu}$, $\tau\sim N^{\hat{\alpha}}$, and $\tau\sim (N/g)^{S_\alpha}$.
The HMF results for $\hat{\nu}$ and $\hat{\alpha}$ are also indicated while
the exponents $S_\nu=S_\alpha=1/2$ are independent of the degree exponent.
 The size of each population is $L=250$ and the   migration rate is $\mu=0.1$. }
\label{tab:exp}
\end{center}
\end{table*}

\begin{figure}[ht]
 \centering
 \includegraphics[width=8.0cm]{\FigPath/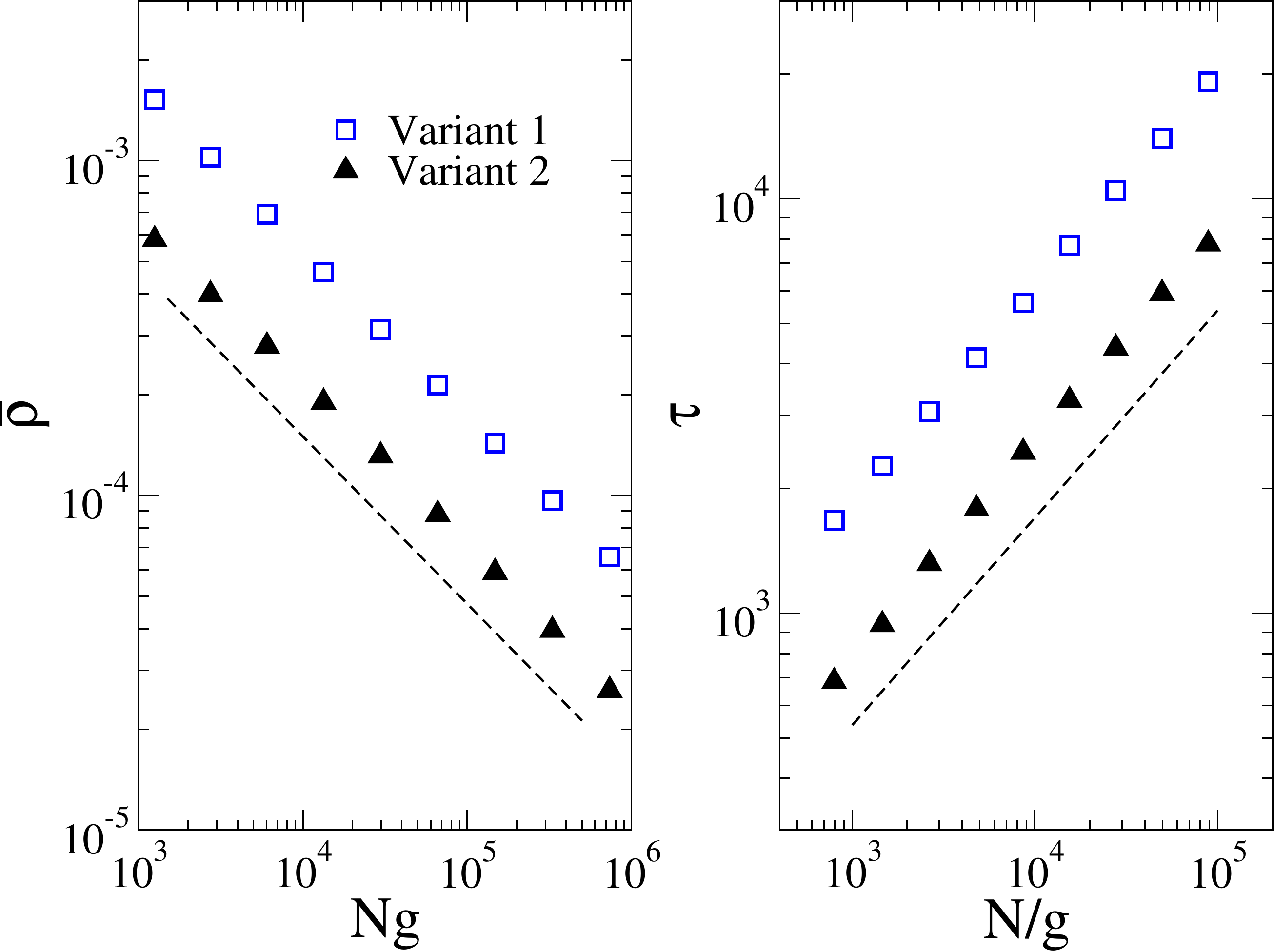}

 \caption{Finite size scaling of the critical density
$\bar{\rho}$ and characteristic time $\tau$ in the variants of the RD
process. In both variants, the structured local 
populations are represented by chains with $z=2$ nearest neighbors. The
migration rate is $\mu = 0.1$ and the degree exponent is $\gamma = 2.75$.
Power laws $(Ng)^{-1/2}$ and $(N/g)^{1/2}$ are shown for comparison.}
\label{fig:variants}
\end{figure}

The high accuracy of HMF to determine critical exponents for local
populations composed of regular chains is somehow surprising since the
structure of the chains strongly violates the hypothesis of
homogeneous mixing used in the mean field approach. However, we can
see that the mixing is indirectly caused by the random movements
through populations mediated by diffusion. This effect is
nevertheless not just due to the random choice of the position inside
the neighbor population.  We have modified the model (variant 1) by using a
deterministic migration rule in the choice of the new position inside
a neighbor chain ($j'\equiv j$ in the model implementation previously
described). This rule mimics, in a very simple fashion, the fact that
individuals use to visit places regularly. This modification does not
change the HMF theory. Numerically, simulations result in a small
increase of the threshold (from $\lambda_c \approx 1.103$ to
$\lambda_c\approx 1.114$ for $\mu=0.1$) due to the reduction of the
spreading power; critical exponents, however, are not altered. Simulations 
confirm the theoretical assertion, see Fig.~\ref{fig:variants}.

The model is also robust to other alterations of the dynamical rules. For
example, we have considered a modification (variant 2) consisting in replacing the
exchange by the creation of a new particle in a neighbor population at
rate $\mu$.  At the HMF level, the last term of equation
Eq.~(\ref{eq:rhok1}) vanishes, the Jacobian matrix
becomes $L_{kk'} = -(1-z\lambda)\delta_{kk'}+\mu C_{kk'}$ and the
critical $\lambda_c = (1-\mu)/z$ is directly found,
indicating that
activity survives for any $\lambda$ if $\mu>1$. Simulations confirm
the HMF predictions: the critical point is reduced (to
$\lambda_c\approx 0.975$ for $\mu=0.1$) but critical exponents remain
unchanged.
Figure
\ref{fig:variants} shows the finite size scaling for critical density
$\bar{\rho}$ and characteristic time $\tau$ for both variants.

\section{Conclusions}
\label{sec:conclu}
In summary, we have investigated the effects of structured populations
in reaction-diffusion (RD) processes 
representing a contact
  process (CP) on metapopulations modeled as complex
networks. Extensive numerical simulations on a variety of model
modifications show that the critical behavior of the RD processes is
essentially independent of the local population structure. The
critical exponents observed are well accounted for by an heterogeneous
mean-field theory (HMF), and depend only on the topological properties
of the metapopulation network; HMF theory fails to reproduce
non-universal properties such as the critical point, which can be
however recovered approximately by using a homogeneous pair
approximation. The agreement between simulations and HMF theory is the
more surprising for local populations with the form of a linear chain,
violating even the small-world property \cite{watts98}. The result
presented here represent compelling evidence of the universality of
the metapopulation RD framework to study 
the critial properties of CP-like models  on
networks, and moreover indicate that an \textit{a priori} strong
oversimplification such as neglecting the local structure is indeed
valid, which implies that a simplified setting with local homogeneous
mixing is enough to reproduce processes in realistic settings.

\begin{acknowledgments}
  This project was supported by Brazilian agencies CNPq, FAPEMIG and
  CAPES.  R.P.-S. acknowledges financial support from the Spanish MEC,
  under project No. FIS2010-21781-C02-01; the Junta de Andaluc\'{\i}a,
  under project No. P09-FQM4682; and ICREA Academia, funded by the
  Generalitat de Catalunya.
\end{acknowledgments}

\appendix

\section{Finite size scaling theory}
\label{app:fss}

A mean field theory for the finite size scaling at criticality of the
RD process under consideration can be obtained using the approach
proposed in Ref.~\cite{castellano08}, in which the motion equation for
$\rho$ is mapped in a one-step process. To obtain  a
self-consistent equation for $\rho$, we explicit compute $\langle
\rho_k^ 2 \rangle$ by squaring Eq.~\eqref{eq:rhok3} and keeping terms
up to order $\mathcal{O}(\rho^2)$.  The result for the stationary
state is
$ \langle \rho_k^2 \rangle \simeq \rho^2 g$,
where $g={\langle k^2 \rangle}/{\km^2}$. In the limit of very low
densities, we can substitute $\langle \rho_k^2 \rangle$ into
Eq.~\eqref{eq:drho_onesite} to obtain the mean-field equation
\begin{equation}
 \label{eq:drho_onesite_mf}
\frac{d\rho}{dt} \simeq - \rho + z\lambda\rho \left[ 1-{\rho g}  \right]. 
\end{equation}
The first term represents an annihilation process $n\rightarrow n-1$
while the second one represents a creation process $n\rightarrow
n+1$. According to Ref.~\cite{castellano08}, the one-step process
corresponding to Eq.~\eqref{eq:drho_onesite_mf} is defined by the
transition rates
\begin{equation}
\label{eq:rates}
 \begin{array}{lll}
  W(n-1,n) & = & n \\
  W(n+1,n) & = & \lambda z n \left[ 1-{\rho g}  \right],
 \end{array}
\end{equation}
where $W(n,m)$ represents the transitions from a state with $m$
occupied sites to another state with $n$ occupied sites. The master
equation for a standard one-step process is~\cite{vankampen}
\begin{equation}
  \label{eq:3}
  \dot{P}_n = \sum_{m} W(n,m) P_m(t) - \sum_m W(m,n) P_n(t).
\end{equation}
Substituting Eq.~\eqref{eq:rates}, we find
\begin{equation}
  \label{eq:master1}
  \dot{P}_n = (n+1)P_{n+1}+u_{n-1}P_{n-1}-(n+u_n)P_n
\end{equation}
with $u_n =\lambda n (1-{\rho g})$.  This equation was investigated in
Ref.~\cite{Ferreira11a} within the QS analysis of the contact process
in annealed scale-free networks. Since the solutions of the
equation~\eqref{eq:master1} have already been exhaustively
investigated, we just report the results from
Ref.~\cite{Ferreira11a}. The critical QS distribution of occupied
vertices for large systems has the form
\begin{equation}
\label{eq:scal_func}
 \bar{P}_n= \frac{1}{\sqrt{\Omega/g}}f\left(\frac{n}{\sqrt{\Omega/g}}\right),
\end{equation}
where $f(x)$ is a scaling function and $\Omega=N\cdot L$ is the total
number of vertices of the metapopulation.  It directly follows from
equation~\eqref{eq:scal_func} that the critical QS density scales as
\[\bar{\rho}  \equiv \frac{1}{\Omega} \sum_n n\bar{P}_n \sim (g\Omega)^{-1/2}\] 
and the characteristic time scales as 
\[\tau \equiv \frac{1}{\bar{P}_1}\sim \left(\frac{\Omega}{g}\right)^{1/2}.\]
For asymptotically large systems with a power law distribution
$P(k)\sim k^{-\gamma}$ we have $g\sim k_c^{3-\gamma}$ for $\gamma<3$
and $g\sim\mbox{const.}$ for $\gamma>3$~\cite{boguna09}. Since $g$
depends only on $N$, the scaling laws $\bar{\rho}\sim
L^{-1/2}N^{-\hat{\nu}}$ and $\tau\sim L^{1/2}N^{\hat{\alpha}}$ ensue
in this way, with critical exponents $\hat{\nu}$ and $\hat{\alpha}$ as
given by Eq.~\eqref{eq:exps} .

\section{Homogeneous pair approximation}
\label{app:PA}
We can develop a homogeneous pair approximation for our RD process
following the procedure described in
Refs.~\cite{Munoz10,sander_phase_2013}.  To do so, let us define the
probability $Q(\sigma\sigma')$ that a pair of neighbors sites in a
homogeneous population are in the states $(\sigma\sigma')$, where
$\sigma=0$ (1) stands for an empty (occupied) node.  The homogeneity
inside populations implies $Q(\sigma\sigma')=Q(\sigma'\sigma)$. Let us
define also the probabilities $\rho=Q(1)$, $\phi=Q(01)$, $\psi=Q(11)$
and $\omega=Q(00)$. The normalization condition $2 \phi+\psi+\omega=1$
and the relation $\rho=\phi+\psi$ apply. The motion equation for
$\rho$ and $\psi$ are
\begin{equation}
\label{eq:drho_pa}
 \frac{d\rho}{dt} = -\rho+z\lambda\phi,
\end{equation}
and
\begin{equation}
\label{eq:psi_pa}
 \frac{d\psi}{dt}  =  -2\psi+ 2\lambda \phi + 2\lambda(z-1)Q(101) +
 2\phi\alpha\rho - 2\psi \alpha(1-\rho). 
\end{equation}
The meaning of the different terms is analogous to those given in the
paper. Equation~\eqref{eq:drho_pa} 
can be directly obtained from Eq.~(\ref{eq:rhok1}) by considering the 
homogeneous approximation, $\rho_k=\rho$ and
$k=\km$, and replacing $\rho(1-\rho)$ by  $\phi$ in the second term.
In a homogeneous pair approximation the probability of a
microscopic configuration is factorized as~\cite{Avraham92}:
\begin{equation}
Q(\sigma_i\sigma_j\sigma_l) \approx
\frac{Q(\sigma_i\sigma_j)Q(\sigma_j\sigma_l)}{Q(\sigma_j)}. 
\end{equation}
Therefore, equation~\eqref{eq:psi_pa} turns to
\begin{equation}
\label{eq:psi_pa2}
 \frac{d\psi}{dt} = -2\psi + 2\lambda\phi + \frac{2(z-1)\lambda\phi^{2}}{(1-\rho)} + 
 2\mu\phi\rho -2\mu\psi(1-\rho). 
\end{equation}
Taking finally the stationary solutions of
equations~\eqref{eq:psi_pa2} and~\eqref{eq:drho_pa}, and using $\rho =
\phi+\psi$, we find the critical point
\begin{equation}
 \lambda_c^{PA} = \frac{1+\mu}{z-1+\mu z}.
\end{equation}

\end{document}